\documentstyle[11pt,newpasp,twoside,epsf]{article}
\def\edcomment#1{\iffalse\marginpar{\raggedright\sl#1\/}\else\relax\fi}
\reversemarginpar

\begin{document}
\title{The ``Number~--~Flux Density'' Relation for Milliarcsecond 
Structures in Extragalactic Radio Sources}
 \author{L.I.~Gurvits}
\affil{Joint Institute for VLBI in Europe \\
P.O.~Box 2, 7990 AA, Dwingeloo, The Netherlands}

\begin{abstract}
``Number~--~flux density'' distributions for total and correlated flux
densities at various interferometer spacings in 355 extragalactic radio
sources are analysed. Qualitative conclusions on the population of
milliarcsecond components in these sources are presented.
\end{abstract}

The progress of Very Long Baseline Interferometry (VLBI) technology
over the past decades made it possible to proceed from studying several
``famous'' sources (as in early VLBI days) to investigating
milliarcsecond radio structures in thousands of extragalactic sources.
In turn, this allows us to study statistical properties of compact
radio structures which bear imprints of both intrinsic source
properties (their evolution in particular) and manifestation of the
cosmological model in the properties of the class of compact radio
sources. One of possible ways of investigating such the statistics is
based on the well established ``number~--~flux density'' apparatus
widely used for studies of extragalactic sources (e.g. Condon 1988).

An illustration of the ``number~--~flux density'' relation for
milliarcsecond radio structures in AGN is based on the data from the
VSOP/VLBA Pre-launch Survey at 5 GHz (VLBApls, Fomalont et al. 2000). In
this survey, 374 extragalactic radio sources, mostly quasars, were
observed with interferometer spacing of up to 150~M$\lambda$
(corresponds to a synthesized beam of about 2~mas). The sample included
all extragalactic sources with a total flux density at 5~GHz, $S_{\rm
5} \ga 1.0$~Jy; a spectral index $\alpha \ge -0.5$~($S \propto
\nu^{\alpha}$); and a galactic latitude $|b| \ge 10\deg$. In addition,
the sample contained all extragalactic sources with  $S_{\rm 5} \ge
5.0$~Jy regardless of their spectral index and galactic coordinates.
See Fomalont et al. (2000) for more discussion on the sample
selection.

Of all the sources observed, 355 were detected. Fig.~1 shows the number
count as a function of total ($S_{\rm tot}$) and correlated flux
densities at 50 and 100~M$\lambda$ ($S_{\rm 50}$ and $S_{\rm 100}$),
respectively. The values of $S_{\rm 50}$ and $S_{\rm 100}$ were
calculated from the selfcalibrated VLBI data as means over the range of
projected baselines [45,55]~M$\lambda$ and [90,110]~M$\lambda$,
respectively. The entire range of flux density values, from 0 to $\sim
50$~Jy, was then divided into 20 nearly equally spaced (in logarithmic
scale) bins. Fig.~1 represents the source counts in the 16 bins with
non-zero number of sources for each of the three flux density values,
$S_{\rm tot}$, $S_{\rm 50}$ and $S_{\rm 100}$.

The oscillations of the $S_{\rm tot}$, $S_{\rm 50}$ and $S_{\rm 100}$
distributions at high ($\ga 4$~Jy) and low ($\la 0.1$~Jy) flux
densities are due to the insufficient statistics (single digit number
of sources in these high and low flux density bins) and is not
significant.

The medium range of flux densities ($0.2 \la S \la 3$~Jy) enables us to
draw the following qualitative conclusions: 

$\bullet$ In the range of $1 \la S \la 3$~Jy, the $S_{\rm 50}$ and
$S_{\rm 100}$ distributions practically coincide, while the $S_{\rm
tot}$ distribution shows roughly three times larger number of sources
in each bin. This could indicate that the compact structures sampled at
50 and 100~M$\lambda$ correspond to the same population of radio
components, distinctive from ``whole'' sources represented by $S_{\rm
tot}$.

$\bullet$ The sharp drop in $S_{\rm tot}$ distribution at $S \la 1$~Jy
is an obvious selection effect. Qualitatively, the $S_{\rm 50}$ and
$S_{\rm 100}$ distributions behave in this range of flux densities as
expected, with the peak of $S_{\rm 100}$ curve at somewhat lower
$S$ than that of $S_{\rm 50}$. If proved significant, the flatter slope
of $S_{\rm 100}$ at $S \la 1$~Jy might indicate a sub-population of
extremely compact and weak ($S_{\rm 100} \la 0.3$~Jy) components.
However, the apparent small difference in the steepness of $S_{\rm 50}$
and $S_{\rm 100}$ distributions at $S \la 1$~Jy indicate that the
ultracompact components do not dominate the milliarcsecond
radio structures in the sample of objects studied.

\begin{figure}
\plotfiddle{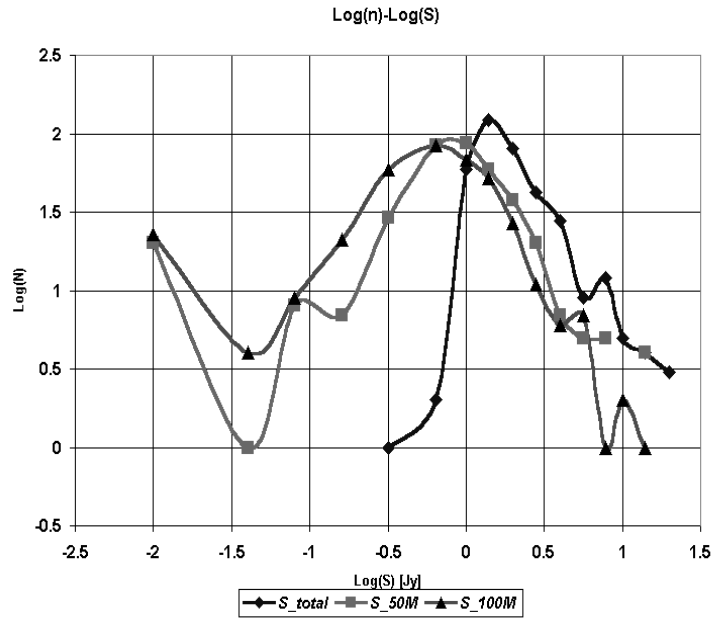}{7.9cm}{0}{82.72}{90}{-135}{0}
\caption{``Number~--~flux~density'' relations for total ($S_{\rm tot}$,
diamond-shaped markers), and correlated flux densities at 50~M$\lambda$
($S_{\rm 50}$, squares) and 100 ~M$\lambda$ ($S_{\rm 100}$, triangles)
in 355 extragalactic sources. }
\end{figure}

{\it Acknowledgment.}~The author is grateful to the Leids Kerkhoven
Bosscha Fonds for travel grant supporting participation in the XXIV IAU
General Assembly and related Symposia.

\end{document}